# Transmission statistics and focusing in single disordered samples

**Matthieu Davy,[1] Zhou Shi,[1,2,*] Jing Wang,[1,2] and Azriel Z. Genack[1,2]**

[1]*Department of Physics, Queens College of the City University of New York, Flushing, NY 11367, USA*
[2]*Graduate Center, City University of New York, New York, NY 10016, USA*
[*]*zhou.shi@qc.cuny.edu*

**Abstract:** We show in microwave experiments and random matrix calculations that in samples with a large number of channels the statistics of transmission for different incident channels relative to the average transmission is determined by a single parameter, the participation number of the eigenvalues of the transmission matrix, $M$. Its inverse, $M^{-1}$, is equal to the variance of relative total transmission of the sample, while the contrast in maximal focusing is equal to $M$. The distribution of relative total transmission changes from Gaussian to negative exponential over the range in which $M^{-1}$ changes from 0 to 1. This provides a framework for transmission and imaging in single samples.

________________________________________________________________

________________________________________________________________

## 1. Introduction

Because of large fluctuations in conductance [1-6] and transmission [7-16] in random systems, transport in individual samples has seemed to be of accidental rather than fundamental significance. Thus, despite wide-ranging applications in communications, imaging and focusing in single samples or environments [17-24], studies of disordered systems have centered on the statistics in hypothesized ensembles of statistically equivalent samples. In contrast to in-depth studies of random ensembles, critical aspects of the statistics of transmission in single samples have not been explored.

Studies of transport have shown that the statistics of propagation over a random ensemble of mesoscopic samples may be characterized in terms of a single parameter, $g$ [25], the ensemble average of the conductance in units of the quantum of conductance $e^2/h$. The wave in random mesoscopic samples is multiply-scattered but temporally coherent throughout the sample [4,5,12]. This is achieved in micron-sized conductors at ultralow temperatures at which electron-phonon interactions are frozen out to the extent that the transit time through the sample is smaller than the electron dephasing time [4,5]. For classical waves, the wavelength of the exciting wave is long and dimensions of scattering elements within the sample are correspondingly large so that thermal motion is much smaller than a wavelength. As a result the transmitted wave remains temporally coherent and stable speckle patterns are observed [7,26]. The similarity of key aspects of quantum and classical wave transport is seen in the equivalence of the dimensionless conductance and the transmittance $T$, which is the sum of flux transmission coefficients between the $N$ incident and transmitted channels, $a$ and $b$, respectively, $g = \langle T \rangle = \langle \sum_{a,b}^{N} T_{ba} \rangle$ [27]. The threshold of the Anderson transition [28] between freely diffusing and spatially localized waves in disordered media lies at $g=1$ [25].

Large sample-to-sample fluctuations are a characteristic feature of transport in mesoscopic systems [1-16]. Conductance and transmission fluctuations relative to the corresponding average over a random ensemble increase exponentially with sample length for localized waves. Even for diffusive waves, the conductance does not self-average as a result of correlation of current in mesoscopic samples [9,10]. In the diffusive limit, $g \gg 1$, conductance fluctuations are universal and the variance of conductance approaches a universal value of order unity [4,5].

We treat the quasi-1D geometry for which the length of reflecting sides greatly exceeds the sample width. The powerful methods of random matrix theory [29-31] were developed in the context of the quasi-1D geometry. Because the wave is completely mixed in the quasi-1D geometry, the statistics of the intensity relative to the average over the transmitted speckle pattern are independent of source or detector positions. Examples of quasi-1D samples are disordered wires and random waveguides, for which measurement are presented here. Thouless showed that electrons become localized when the length of a wire exceeds the localization length $\xi = N\ell / L$, where $\ell$ is the mean free path and $L$ is the sample length [32]. Beyond this length the resistance increases exponentially. In direct analogy to the predicted localization of electrons in wires, a crossover to localization has been observed for microwave radiation in ensembles of random waveguides of increasing length but constant cross section [33]. Because of the Gaussian distribution of the field in any single speckle pattern, the probability distribution of $T_{ba} / (\sum_{b}^{N} T_{ba} / N) = NT_{ba} / T_a$ relative intensity is a negative exponential, $P(NT_{ba} / T_a) = \exp(-NT_{ba} / T_a)$ [12,34]. Since the statistics of relative intensity are universal, the statistics of transmission in a sample with transmittance $T$ would be completely specified by the statistics of total transmission $T_a$ relative to its average $T/N$ within the sample.

We report here the essential statistics of transmission in single transmission matrices as opposed to the statistics of random ensembles. We find the statistics of relative total transmission $NT_a / N$ and show it is determined by a single parameter, the participation number of the eigenvalues $\tau_n$ of the matrix product $tt^{\dagger}$, $M = (\sum_n \tau_n)^2 / \sum_n \tau_n^2$ [23], where $t$ is the transmission matrix connects the incident field and outgoing field. The sum over all $\tau_n$ in a given matrix is equal to the transmittance $T = \sum_{n=1}^{N} \tau_n$ [29-31]. We demonstrate using measurements of the microwave transmission matrix and random matrix calculations that, in the limit of large $N$, $M^{-1}$ is equal to the variance of $NT_a / T$. The distribution of relative total transmission changes from Gaussian to negative exponential over the range in which $M^{-1}$ changes from 0 to 1. The contrast in maximal focusing [18] is found as a function of $M$ and $N$. In the limit of large $N$, the contrast is equal to $M$.

## 2. Statistics of single transmission matrices

The microwave transmission matrix $t$ is recorded for random ensembles with values of $g$ ranging from 6.9 to 0.17. The random samples are mixtures of 0.95-cm-diameter alumina spheres of refractive index $n=3.14$ embedded within Styrofoam shells to give an alumina volume fraction of 0.069 within the 7.3-cm diameter sample tube. Polarized microwave radiation is produced by a dipole antenna connected to the network analyzer and the transmitted electric field is measured with a 4-mm wire antenna. The polarization is determined by the orientation of the antenna. Both antennas are mounted on a two-dimension translation stage so that they can move freely on the input and output surfaces. The transmission matrix is recorded over a grid of $N/2$ points on the input and output of the sample. The antennas are also rotated between two orthogonal orientations to measure $N^2$ elements of $t$, in which $N$ is the number of propagation modes allowed in the waveguide. Measurements are analyzed for samples of lengths $L$=23, 40 and 61 cm in two frequency ranges 14.7-14.94 GHz and 10-10.24 GHz in which the wave is diffusive and localized, respectively, with $N$~66 and 30. New configurations were obtained by momentarily rotating the copper tube about its axis once the full matrix is recorded. The impact of absorption on the statistics of transmission is removed by Fourier transforming the field spectrum into the time domain and then multiplying the time signal by $e^{t/2\tau_a}$, where $t$ is the time delay and $1/\tau_a$ is the energy absorption rate. The time dependent signal is then Fourier transformed back to the frequency domain [13].

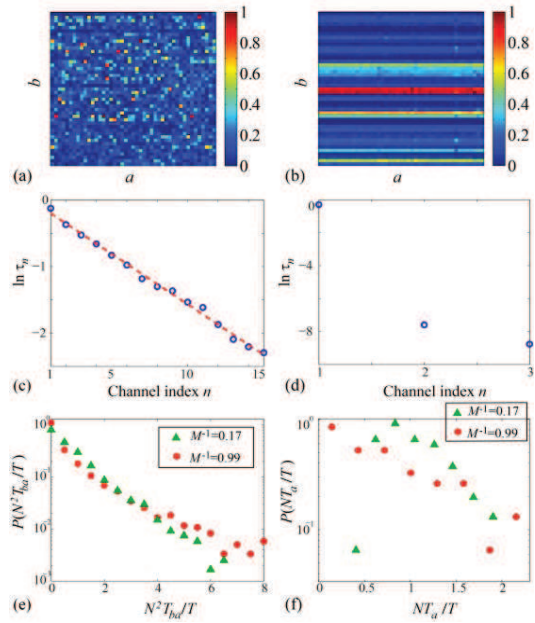

Fig. 1. Intensity normalized to the peak value in each speckle pattern generated by sources at positions $a$ are represented in the columns with index of detector position $b$ for (a) diffusive and (b) localized waves. (c,d) The transmission eigenvalues are plotted under the corresponding intensity patterns. For localized waves (d), the determination of the third eigenvalue and higher eigenvalues are influenced by the noise level of the measurements. Correlation between speckle patterns for different source positions are clearly seen in (b) due to the small numbers of eigenchannels $M$ contributing appreciably to transmission. (e,f) Distributions of relative intensity P($N^2T_{ba}/T$) and relative total transmission P($NT_a/T$) for the two transmission matrices selected in this figure with $M^{-1}$=0.17 (green triangles) and $M^{-1}$=0.99 (red circles).

In Figs. 1(a) and 1(b), we show intensity pattern within a single transmission matrix drawn from ensembles with $g$=6.9 and 0.17, respectively. Plots of the transmission eigenvalues $\tau_n$ determined from the transmission matrices for the samples whose intensity patterns are shown in Figs. 1(a) and 1(b) are presented below the corresponding patterns. Values of $\tau_n$ are seen to be substantial for a number of channels and to fall nearly exponentially for the sample in which the wave is diffusive. Since the transmitted speckle pattern for a source at any position $a$ is the sum of many orthogonal transmission eigenchannels, speckle patterns for different source positions are weakly correlated yielding the motley intensity pattern for the transmission matrix depicted in Fig. 1(a). In contrast, the first transmission eigenchannel for localized waves dominates transmission in Fig. 1(d) so that the normalized speckle patterns for each input are highly correlated and horizontal stripes appear in the Fig. 1(b). The probability distributions of relative intensity, $N^2 T_{ba}/T$, and total transmission, $NT_a/T$, for the two transmission matrices depicted in Fig. 1(a) and 1(b) are shown in Figs. 1(e) and 1(f).

Previous findings that propagation in random ensembles can be characterized via the variance of the total transmission relative to its ensemble average, var($NT_a$/<$T$>) [12,13] suggest that the statistics of single samples may be characterized via the variance of relative total transmission in a single transmission matrix, var($NT_a/T$). This can be calculated for a single instance of the transmission matrix with $N \gg 1$ using the singular value decomposition of the transmission matrix, $t = U \Lambda V^\dagger$. Here, $U$ and $V$ are unitary matrices with elements $u_{nb}$ and $v_{na}$, where $n$ is the index of the eigenchannel and index $a$ and $b$ indicate the input and output channels, respectively. The real and imaginary parts of $v_{na}$ are Gaussian random variables with zero mean and variance of $1/2N$. $\Lambda$ is a diagonal matrix with elements $\sqrt{\tau_n}$.

The total transmission from incident channel $a$, $T_a = \sum_b T_{ba}$, can be written as, $T_a = \sum_{n=1}^{N} \tau_n |v_{na}|^2$, giving the relative total transmission, $T_a/(T/N) = \sum_{n=1}^{N} \tau_n \left( N |v_{na}|^2 \right)/T$. The second moment of $NT_a/T$ is,

$$\sum_{n=1}^{N} (\tau_n/T)^2 \left\langle N^2 |v_{na}|^4 \right\rangle_a + \sum_{n' \neq n}^{N} \tau_n \tau_{n'}/T^2 \left\langle N^2 |v_{na}|^2 |v_{n'a}|^2 \right\rangle_a. \qquad (1)$$

Here, $\langle \cdots \rangle_a$ is the average over all the incident points $a$ within a single transmission matrix. In the limit $N \gg 1$, $\langle |v_{an}|^2 \rangle = 1/N$ and $\langle |v_{na}|^4 \rangle = 2/N^2$. This yields,

$$\left\langle \left( \frac{NT_a}{T} \right)^2 \right\rangle_a = \frac{\sum_{n=1}^{N} \tau_n^2 + \left( \sum_{n=1}^{N} \tau_n \right)^2}{\left( \sum_{n=1}^{N} \tau_n \right)^2}. \qquad (2)$$

Since $\langle NT_a/T \rangle_a = 1$ in a given transmission matirx, this gives

$$\text{var}(NT_a/T) = \sum_{n=1}^{N} \tau_n^2 / \left( \sum_{n=1}^{N} \tau_n \right)^2 \equiv M^{-1}. \qquad (3)$$

We had previously denoted the eigenvalues participation number, $M$, by $N_{\text{eff}}$ [23], but changed the notation here since this parameter is distinct from $N_{\text{eff}}$, which had been introduced earlier by Imry [35] to denote the number of channels with $\tau_n > 1/e$. Whereas $N_{\text{eff}} \to 0$ in the localization limit, $N_{\text{eff}} \to 0$ in this limit since transmission is then dominated by a single eigenchannel.

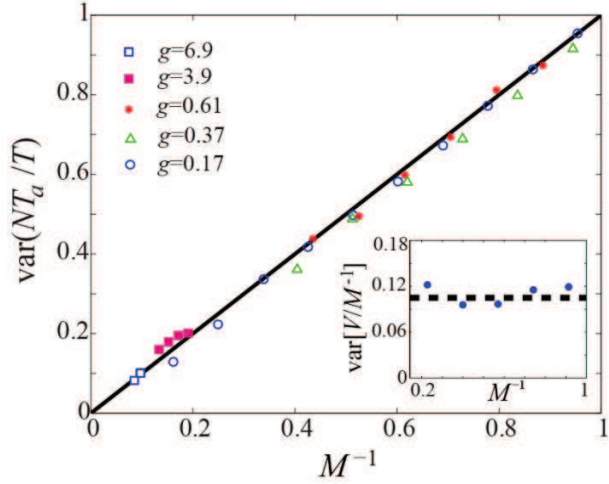

Fig. 2. Plot of the var($NT_a/T$) computed within transmission matrices over a subset of transmission matrices drawn from random ensembles with different values of $g$ with specified value of $M^{-1}$. The straight line is a plot of var($NT_a/T$)=$M^{-1}$. In the inset, the variance of V/$M^{-1}$ is plotted vs. $M^{-1}$, where V=var($NT_a/T$).

These calculations are confirmed in a comparison with measurements in samples of small $N$ which are binned together with samples with similar values of $M^{-1}$. The average of var($NT_a/T$) in subsets of samples with given $M^{-1}$ is seen in Fig. 2 to be in excellent agreement with Eq. (1). var[var($NT_a/T$)/$M^{-1}$] is seen in the insert of Fig. 2 to be proportional to 1/$N$ indicating that fluctuations in the variance over different subsets are Gaussian with a variance that vanishes as $N$ increases.

The central role played by $M$ can be appreciated from the plots shown in Fig. 3 of the statistics for subsets of samples with identical values of $M$ but drawn from ensembles with different values of $g$. The distributions $P(NT_a/T)$ obtained for samples with $M^{-1}$ in the range 0.17±0.01 selected from ensembles with $g$=3.9 and 0.17 are seen to coincide in Fig. 3(a) and thus to depend only on $M^{-1}$. The curve in Fig 3(a) is obtained from an expression for $P(T_a/\langle T_a\rangle)$ for diffusive waves given in Ref. [12], in terms of the single parameter $g$, which equals 2/3var($T_a/\langle T_a\rangle$) in the limit of large $g$, in which $g$ is replaced by 2/3$M^{-1}$. The dependence of $P(NT_a/T)$ on $M^{-1}$ alone and its independence of $T$ is also demonstrated in Fig. 3(b) for $M^{-1}$ over the range 0.995±0.005 in measurements in different sample length with $g$=0.37 and 0.17. Since a single channel dominates transmission in the limit, $M^{-1} \to 1$, we have, $NT_a/T = |v_{1a}|^2$. The Gaussian distribution of the elements of $V$ leads to a negative exponential distribution for the square amplitude of these elements and similarly to $P(NT_a/T) = \exp(-NT_a/T)$, which is the curve plotted in Fig. 3(b). In Figs. 3(c) and 3(d), we plot the relative intensity distributions $P(N^2T_{ba}/T)$ corresponding to the same collection of samples as in Figs. 3(a) and 3(b), respectively. The curves plotted are the intensity distributions obtained by mixing the distributions $P(NT_a/T)$ shown in Figs. 3(a) and 3(b) with the universal negative exponential function for $P(NT_{ba}/T_a)$.

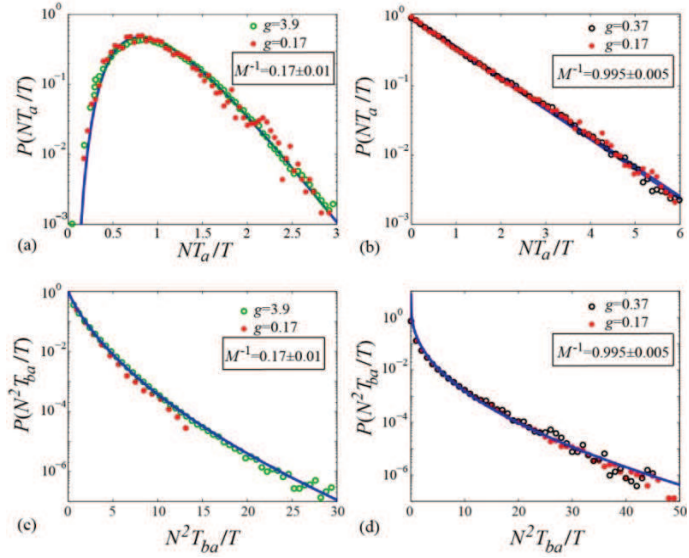

Fig. 3. (a) $P(NT_a/T)$ for subsets of transmission matrices with $M^{-1}=0.17\pm0.01$ drawn from ensembles of samples with $L=61$ cm in two frequency ranges in which the wave is diffusive (green circles) and localized (red filled circles). The curve is the theoretical probability distribution of $P(NT_a/<T>)$ in which var$(NT_a/T)$ is replaced by $M^{-1}$ in the expression for $P(NT_a/T)$ in Ref. 12. (b) $P(NT_a/T)$ for $M^{-1}$ in the range $0.995\pm0.005$ computed for localized waves in samples of two lengths: $L=40$ cm (black circles) and $L=61$ cm (red filled circles). The straight line represents the exponential distribution, $\exp(-NT_a/T)$. (c,d) The corresponding intensity distributions $P(N^2T_{ba}/T)$ are plotted under (a) and (b).

The departure of intensity and total transmission distributions within a transmission matrix from negative exponential and Gaussian distributions, respectively, is a consequence of mesoscopic intensity correlation. The results above for the statistics over a subset of samples with given $M$ suggest an expression for the cumulant correlation function of transmitted intensity relative to its average value for a single transmission matrix in the limit $N\gg 1$ or for a subset of transmission matrices with a specified value of $M$,

$$C^M_{ba,b'a'} = \langle [T_{ba}T_{b'a'} - (T/N^2)^2]/(T/N^2)^2 \rangle_M . \qquad (4)$$

Because of the normalization by the $T/N^2$, infinite-range correlation of relative intensity between arbitrary incident and outgoing channels for transmission matrices with given $M$ vanishes. Such correlation in a random ensemble is known as $C_3$ and is due to fluctuations in $T$. [9,10] The values of unity and $M^{-1}$ for the variances of relative intensity and total transmission determines the sizes of the residual $C_1$ and $C_2$ terms, representing short- and long-range intensity correlation within the matrix, respectively and gives

$$C^M_{ba,b'a'} = \delta_{aa'}\delta_{bb'} + M^{-1}(\delta_{aa'} + \delta_{bb'}). \qquad (5)$$

This gives $\text{var}(N^2T_{ab}/T) = 1 + 2M^{-1}$. This is confirmed by the close correspondence between the measured variances of relative intensity for the two values of $M$ of 0.17 and 0.995 in Fig. 3 of 1.38 and 3.04 with the calculated values of 1.34 and 3.

## 3. Focusing in single transmission matrices

The statistics of single transmission matrices provide the basis for the control of transmission. A key parameter is the contrast in optimal focusing that is achieved with phase conjugation

[18]. In order to focus wave transmitted through a random medium to a target point $\beta$, one can apply the wavefront $t^*_{\beta a}/\sqrt{T_\beta}$ so that the transmitted electric field from different incident points $a$ arrives at $\beta$ in phase and interferes constructively. Here, the incident field is normalized by $T_\beta = \sum_{a=1}^{N}|t_{\beta a}|^2$ to set the incident power to be unity. With this normalization, the intensity at the focal point is $I_\beta = T_\beta$. The intensity away from the focus spot, can be expressed in terms of elements of the $U$ and $V$ matrices obtained in singular value decomposition,

$$I_{b\neq\beta} = \frac{\left|\sum_{n=1}^{N}\tau_n u_{nb} u^*_{n\beta}\right|^2}{T_\beta} = \frac{\sum_{n,n'=1}^{N}\tau_n \tau_{n'} u_{nb} u^*_{n\beta} u^*_{n'b} u_{n'\beta}}{\sum_{n=1}^{N}\tau_n |u_{n\beta}|^2}. \quad (6)$$

We consider the ratio within a single transmission matrix between the average intensity at the at the and the average intensity in the background as the focusing contrast, $\mu = \langle I_\beta \rangle_\beta / \langle I_{b\neq\beta}\rangle_{b,\beta}$, where $\langle \cdots \rangle_\beta$ indicates averaging over all possible focusing points. This is slightly different from $\langle I_\beta / \langle I_{b\neq\beta}\rangle_b \rangle_\beta$ defined in Ref. 23. The average intensity at the focal point is $\langle I_\beta \rangle_\beta = \sum_{n=1}^{N}\tau_n / N$. The intensity averaged over all points $b \neq \beta$ is,

$$\langle I_b \rangle_{b\neq\beta} = \frac{1}{N-1}\frac{\sum_b \sum_{n,n'=1}^{N}\tau_n \tau_{n'} u_{nb} u^*_{n\beta} u^*_{n'b} u_{n'\beta}}{\sum_{n=1}^{N}\tau_n |u_{n\beta}|^2} - \frac{1}{N-1}\frac{\left(\sum_{n=1}^{N}\tau_n |u_{n\beta}|^2\right)^2}{\sum_{n=1}^{N}\tau_n |u_{n\beta}|^2}. \quad (7)$$

This gives

$$\langle I_b \rangle_{b\neq\beta} = \frac{1}{N-1}\frac{\sum_{n=1}^{N}\tau_n^2 |u_{n\beta}|^2}{\sum_{n=1}^{N}\tau_n |u_{n\beta}|^2} - \frac{1}{N-1}\sum_{n=1}^{N}\tau_n |u_{n\beta}|^2. \quad (8)$$

The background intensity is then averaged over the focusing point $\beta$, which gives the contrast in the limit of $N\gg1$ as,

$$\mu = \frac{1}{1/M - 1/N}. \quad (9)$$

The results of measurements shown in Fig. 4 are in excellent agreement with Eq. (9). When the number of measured points $N'$ is smaller than $N$ and therefore the corresponding eigenvalue participation number $M'$ is smaller than $M$, the contrast is given by Eq. (9) with the substitutions $M \to M'$ and $N \to N'$. This is demonstrated in Fig. 4 with the contrast in samples with $N=66$ but with the contrast computed only for $N'=30$ points falling on the curve for $N=30$. These results may be applied to optical measurements of the transmission matrix in which the size of the measured matrix $N'$ is generally much smaller than $N$. In the limit $N \gg M$, the contrast approaches $M$.

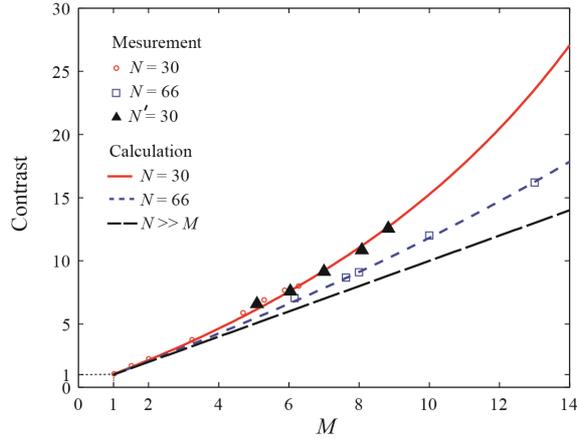

Fig. 4. Contrast in maximal focusing vs. eigenchannel participation number $M$. The open circles and squares represent measurements from transmission matrices $N = 30$ and $66$ channels, respectively. The filled triangles give results for $N'\times N'$ matrices with $N' = 30$ for points selected from a larger matrix with size $N=66$. Phase conjugation is applied within the reduced matrix to achieve maximal focusing. Equation (3) is represented by the solid red and dashed blue curves for $N=30$ and $66$, respectively. In the limit of $N>>M$, the contrast given by Eq. (3) is equal to $M$, which is shown in long-dashed black line.

## 4. Conclusion

We have shown that $M$ and $T$ have distinct roles in individual transmission matrices. $M$ determines the statistics of relative transmission in single matrices while $T$ serves as an overall normalization factor for transmission. In large matrices, the inverse of $M$ equals the variance of the total transmission while $M$ gives the contrast in optimal focusing. In applications in which both relative and absolute transmission play a role, the separate roles of $T$ and $M$ can be seen. For example, in maximal focusing, the peak intensity is equal to $T/N$, while the contrast in the averaged focused pattern is equal to $1/(1/M-1/N)$. In contrast to the need for two parameters to treat the statistics within single transmission matrices, the statistics of a random ensemble depend only upon a single parameter, $g =<T>$. Whereas, $<M>$ is a function of $g$ and is proportional to $g$ for diffusive waves.

### Acknowledgements

We thank Howard Rose for advice on the operation of the experiment. The research was supported by the NSF under Grant No. DMR-1207446 and by the Direction Générale de l'Armement (DGA).